\documentclass[baaa]{baaa}

\usepackage[pdftex]{hyperref}
\usepackage{subfigure}
\usepackage{natbib}
\usepackage{helvet,soul}
\usepackage[font=small]{caption}

\contriblanguage{1}

\contribtype{2}

\thematicarea{1}

\title{Pixel color-magnitude diagrams of galaxies in the \\
  Fornax cluster using S-PLUS images}

\titlerunning{Pixel color-magnitude diagrams}

\author{
  C.G. Escudero\inst{1,2},
  A.V. Smith Castelli\inst{1,2}
  F.A. Faifer\inst{1,2},
  L.A. Sesto\inst{1,2},
  C.L. Mendes de Oliveira\inst{3},\\
  F.R. Herpich\inst{3}
\&
  C.E. Barbosa\inst{3}
}

\authorrunning{Escudero et al.}

\contact{cgescudero@fcaglp.unlp.edu.ar}

\institute{
Facultad de Ciencias Astronómicas y Geofísicas, UNLP, Argentina
\and
Instituto de Astrofísica de La Plata, CONICET--UNLP, Argentina
\and
Instituto de Astronomía, Geofísica y Ciencias Atmosféricas, USP, Brasil
}

\resumen{
 Los cúmulos de galaxias son los sistemas ligados gravitacionalmente más grandes del Universo y, como tales, juegan un papel importante en los estudios cosmológicos. Un recurso importante para estudiar sus propiedades de manera estadística es utilizar conjuntos de datos de imágenes homogéneos y de gran cobertura areal que puedan abarcar diversos medioambientes. En este sentido, las imágenes de campo amplio (1.4 grados$^2$) obtenidas por el {\em Southern Photometric Local Universe Survey} (S-PLUS) en 12 bandas ópticas, constituyen una valiosa herramienta para ese tipo de estudios. En este trabajo, presentamos un análisis fotométrico de los denominados diagramas color-magnitud pixelados, correspondiente a una muestra de 24 galaxias de diversos tipos morfológicos ubicadas en el cúmulo de Fornax.

}

\abstract{
Galaxy clusters are the largest gravitationally bound systems in the Universe and, as such, play an important role in cosmological studies. An important resource for studying their properties in a statistical manner are homogeneous and large image datasets covering diverse environments. In this sense, the wide-field images (1.4 deg$^2$) obtained by the Southern Photometric Local Universe Survey (S-PLUS) in 12 optical bands, constitute a valuable tool for that type of studies. In this work, we present a photometric analysis of pixel color-magnitude diagrams, corresponding to a sample of 24 galaxies of different morphological types located in the Fornax cluster.
  
}

\keywords{surveys --- galaxies: clusters: individual (Fornax) --- galaxies: photometry}

\begin{document}

\maketitle

\section{Introduction}\label{S_intro}
Galaxy clusters are the largest gravitationally bound systems in the Universe and, as such, play an important role in cosmological studies. According to the hierarchical clustering scenario, galaxy clusters continuously grow by accreting individual galaxies, as well as entire galaxy groups. Therefore, they have been regarded as powerful laboratories for studying the physical processes that might have influenced galaxies evolution, such as strong galaxy-galaxy interaction, ram pressure stripping, among others. All these environmental mechanisms result in different morphological properties for cluster and field galaxies. An important resource for studying these properties in a statistical manner are homogeneous and large image datasets covering diverse environments. In this sense, the wide-field images obtained by the Southern Photometric Local Universe Survey \citep[S-PLUS;][]{deOliveira2019} in 12 optical bands, constitute a valuable tool for that type of study (see Section \ref{sec:obs} for details).

In this work, we present the pixel color-magnitude diagrams \citep[pCMDs; e.g.,][]{Foster2007} of a sample of 24 galaxies of different morphological types located in the Fornax cluster \citep[d=21.1 Mpc;][]{Blakeslee2010}. Our main goal is to elucidate if this kind of diagrams could be used to detect internal substructures in galaxies and if they could help to carry out a morphological classification of galaxies. According to the distance adopted for Fornax, 1 arcmin corresponds to a scale of 5.81 kpc. 


\section{S-PLUS and pCMD}\label{sec:obs}
S-PLUS is an imaging survey that will cover $\sim9300$ deg$^2$ in 12 filters, using a robotic 0.8 m-aperture telescope located at the Cerro Tololo Inter-American Observatory (CTIO), Chile. Its wide field camera allows to obtain images of 1.4 $\times$ 1.4 deg$^2$ with a pixel scale of 0.55 arcsec. In addition to the broadband filters $u,g,r,i,z$, S-PLUS includes 7 narrowband filters centered on the following spectral features: [OII], Ca H+K, H$\delta$, G-band, Mgb triplet, H$\alpha$ and Ca triplet. The characteristics of the survey allow performing an analysis of a large number of objects in a homogeneous manner. 

From these photometric data, we built individual pCMDs for 24 galaxies in the Fornax cluster, where each point in these diagrams corresponds to a pixel in the image. However, previously, it was necessary to homogenize the seeing values in the different images and mask the objects near the galaxies, in order to avoid spurious effects in the diagrams. Since the typical seeing value in these images is between 1-2 arcsec, to obtain a pixel scale of the order of the point spread function (PSF), we considered a binning factor of 2. By doing this, we make sure that each pixel is statistically independent of the surrounding pixels, and we improve the signal-to-noise (S/N) ratio in the data. Finally, the flux of each pixel in the 12 bands was converted to apparent magnitude per square arcsec and calibrated to the standard system. We consider the limit at $g_0=22$ mag\,arcsec$^{-2}$ in the diagrams to ensure an S/N$>$5 in each pixel. In this way, it is possible to reach several arcsec of galactocentric radii before the background level begins to affect the diagrams.


\section{Morphology}\label{sec:morphology}
Observing the different pCMDs, a wide range of shapes and features can be seen. However, grouping the galaxies according to their morphological classification \citet{devaucouleurs1991},
some similarities between the diagrams emerge.

\begin{figure}[!t]
\centering
\includegraphics[width=0.49\columnwidth]{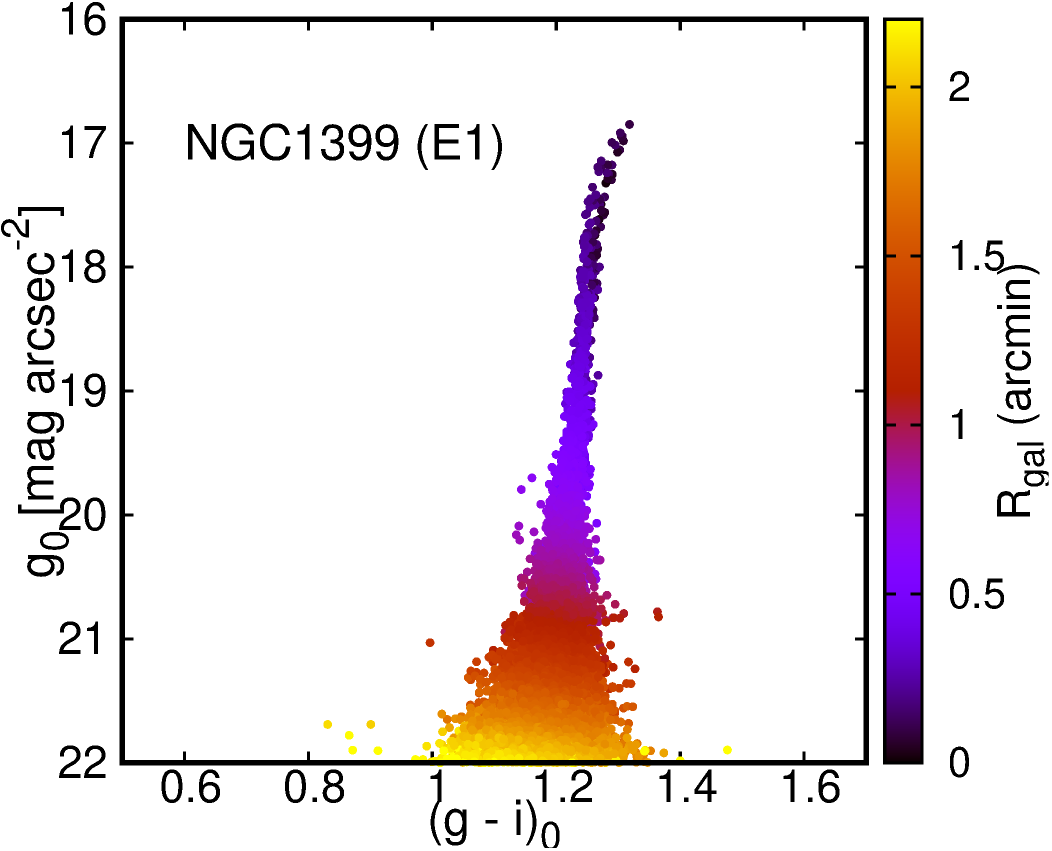}
\includegraphics[width=0.49\columnwidth]{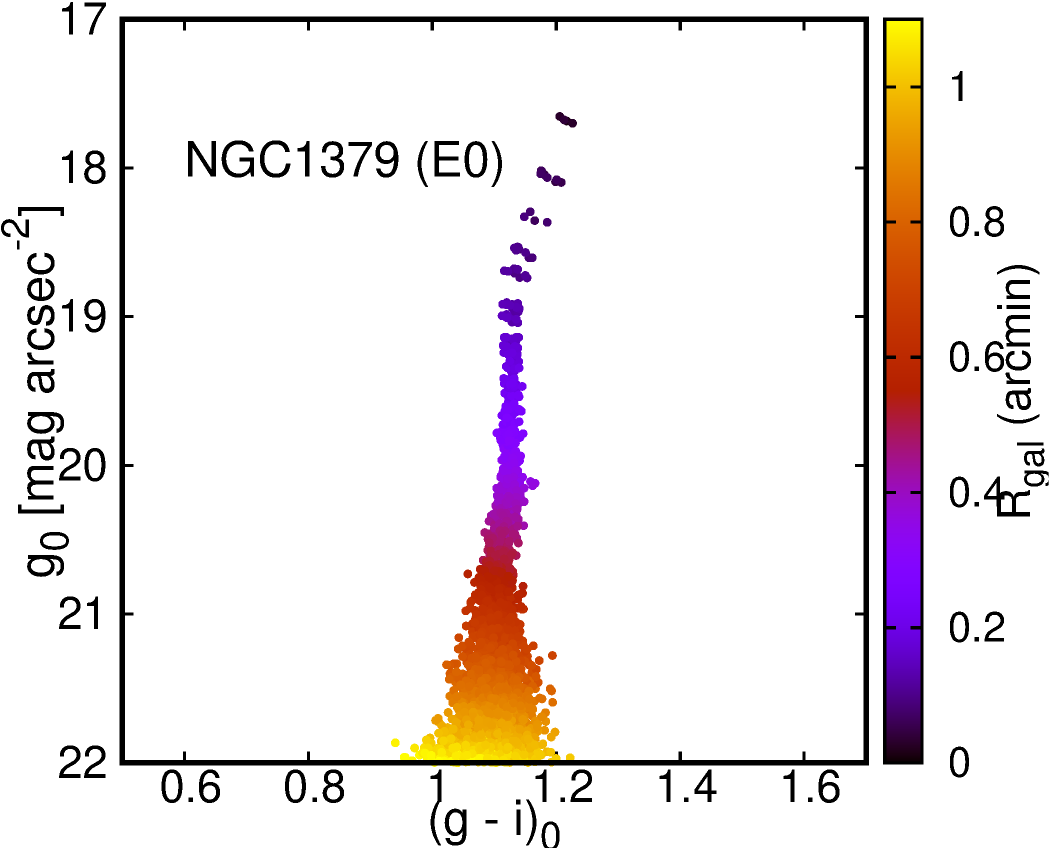}
\includegraphics[width=0.49\columnwidth]{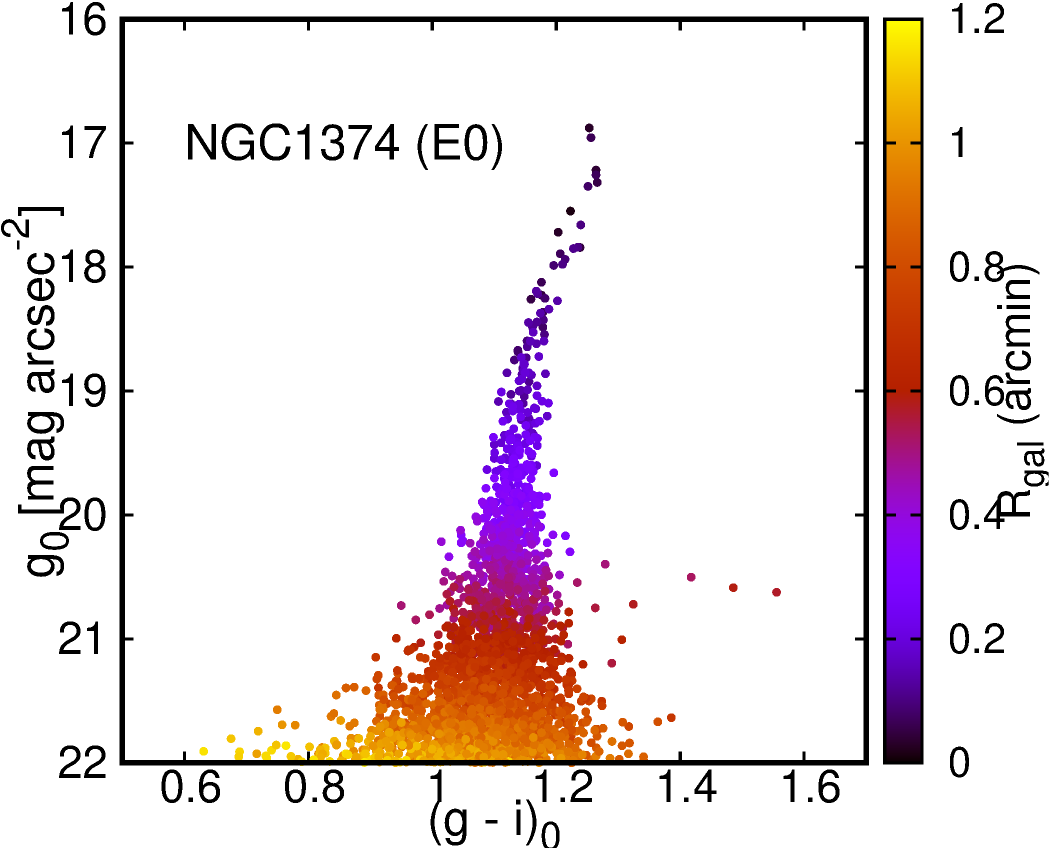}
\includegraphics[width=0.49\columnwidth]{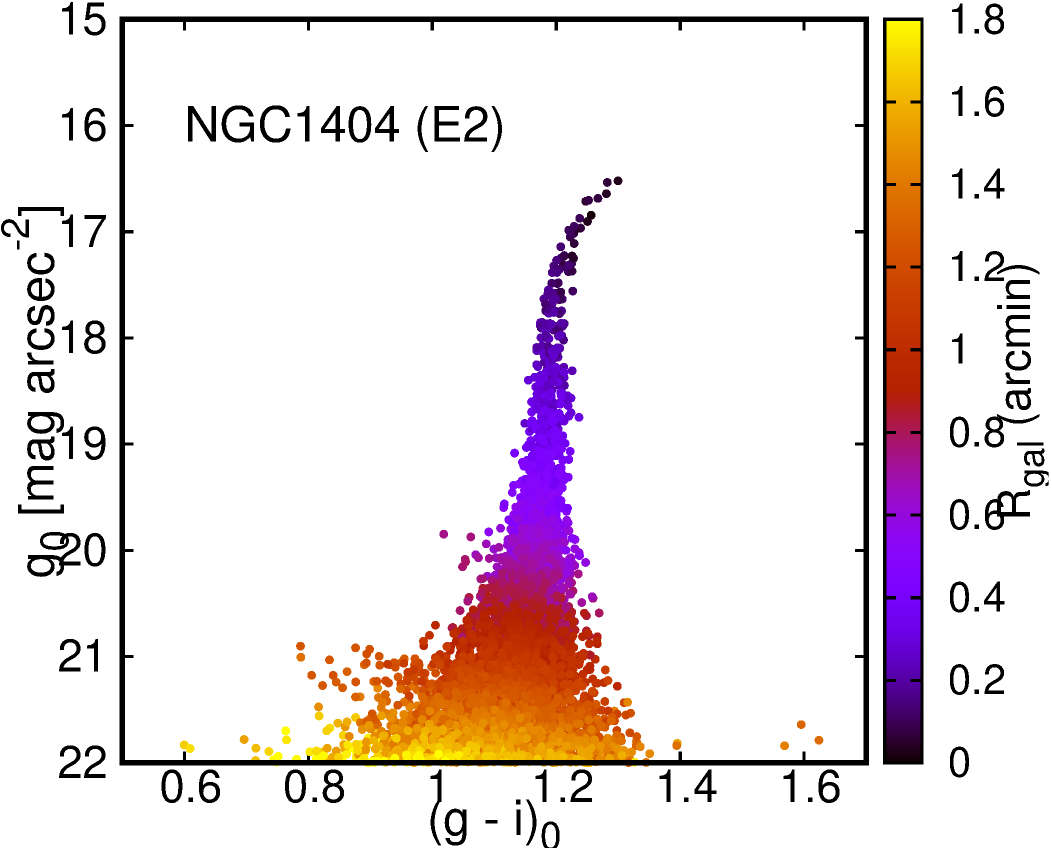}
\caption{Pixel color-magnitude diagrams of the analyzed elliptical galaxies, organized by decreasing luminosity (from left to right, and from top to bottom). The morphological type is indicated between parentheses. The color bar gives the galactocentric distance code for each pixel.
}
\label{Figure1}
\end{figure}

\begin{figure*}[!t]
\centering
\includegraphics[width=0.5\columnwidth]{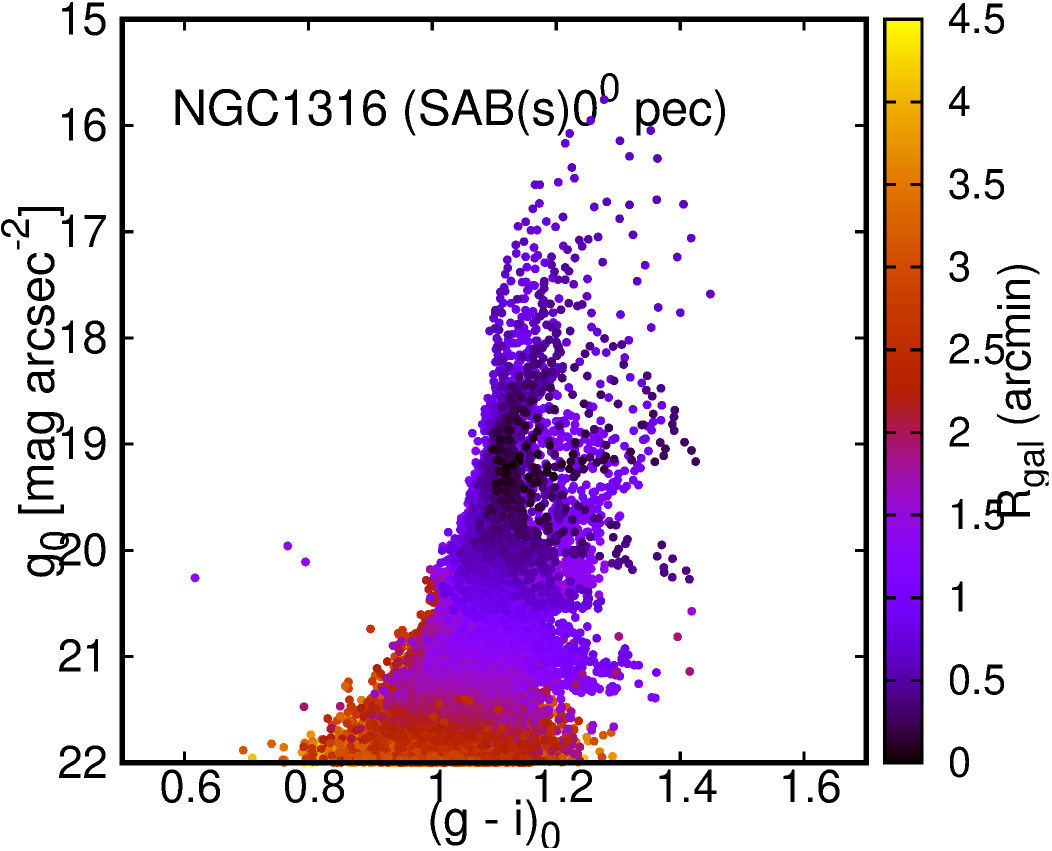}
\includegraphics[width=0.5\columnwidth]{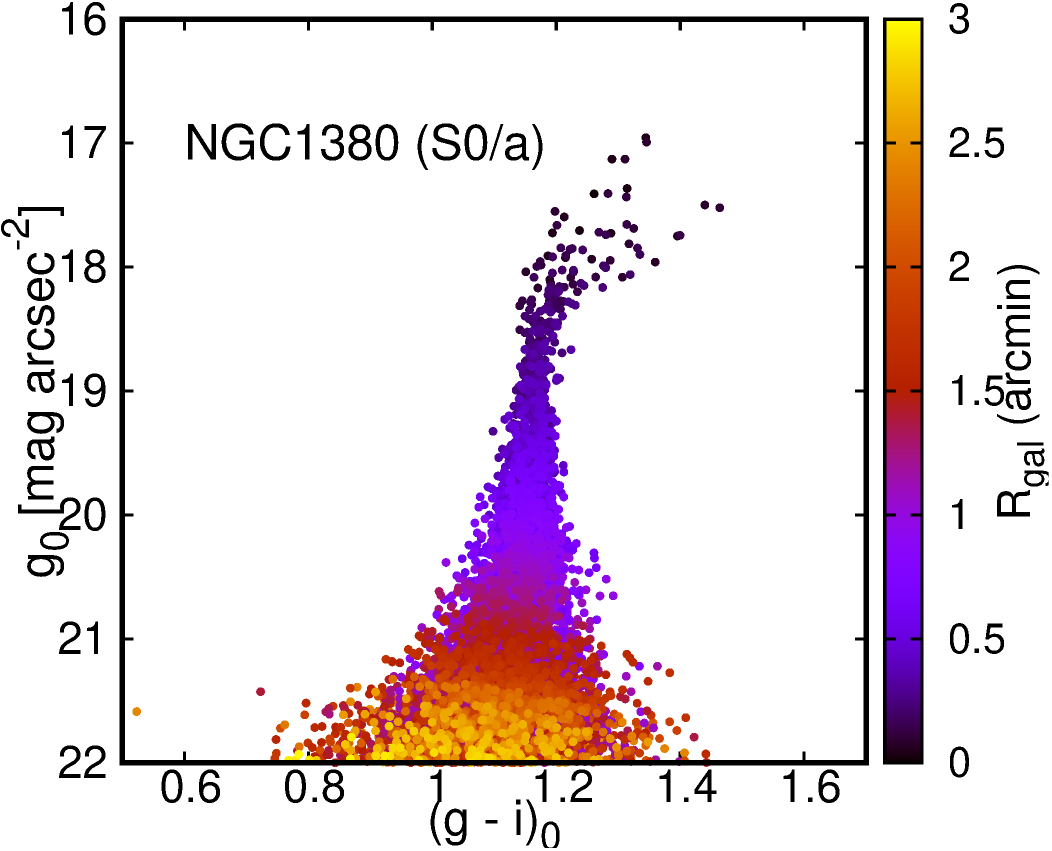}
\includegraphics[width=0.5\columnwidth]{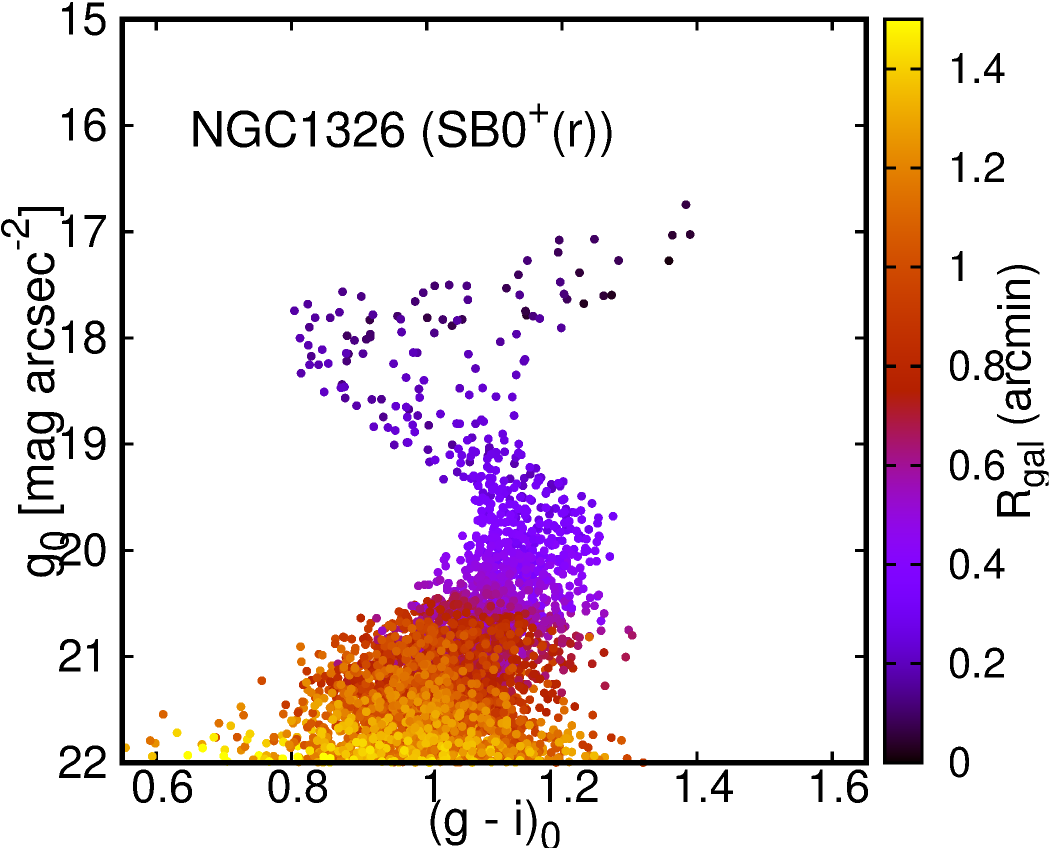}
\includegraphics[width=0.5\columnwidth]{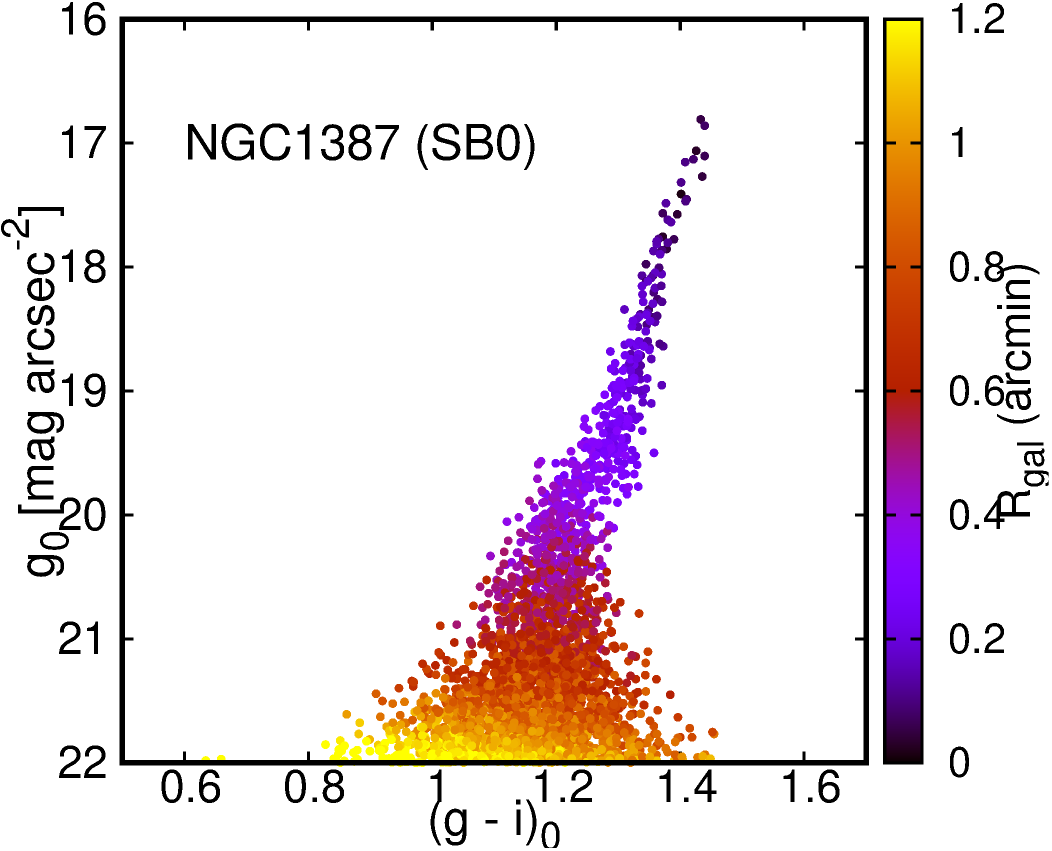}
\includegraphics[width=0.5\columnwidth]{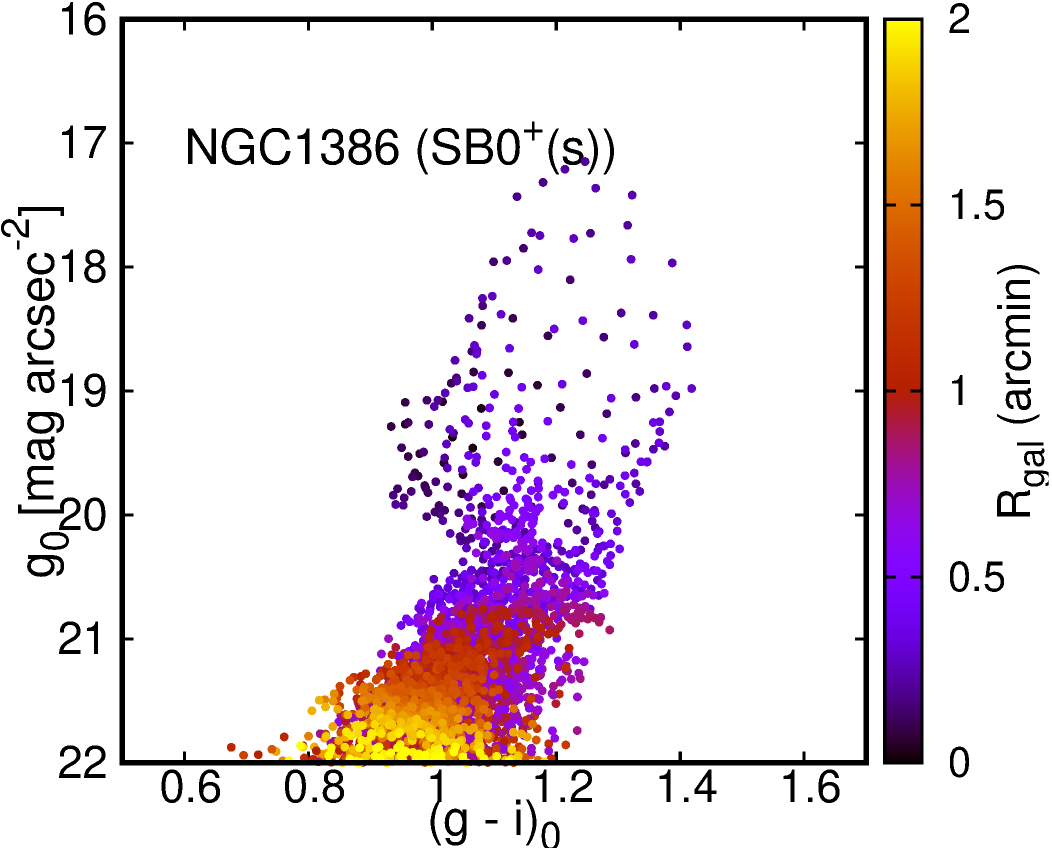}
\includegraphics[width=0.5\columnwidth]{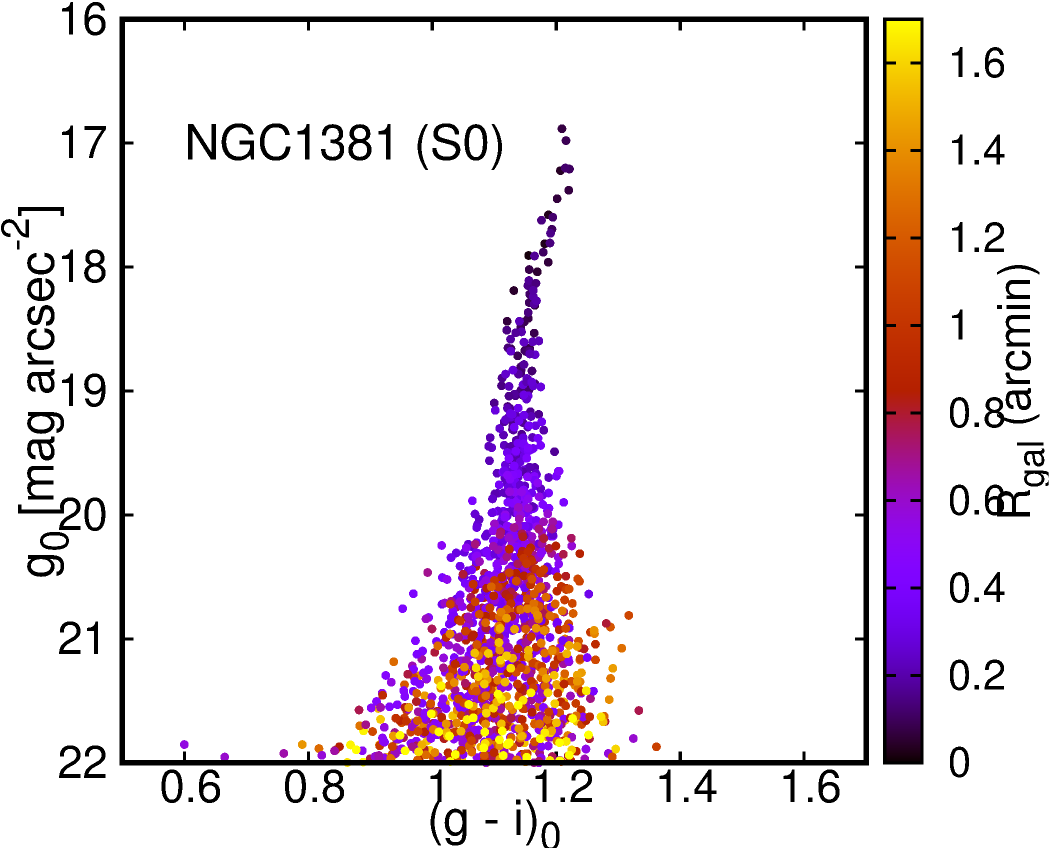}
\includegraphics[width=0.5\columnwidth]{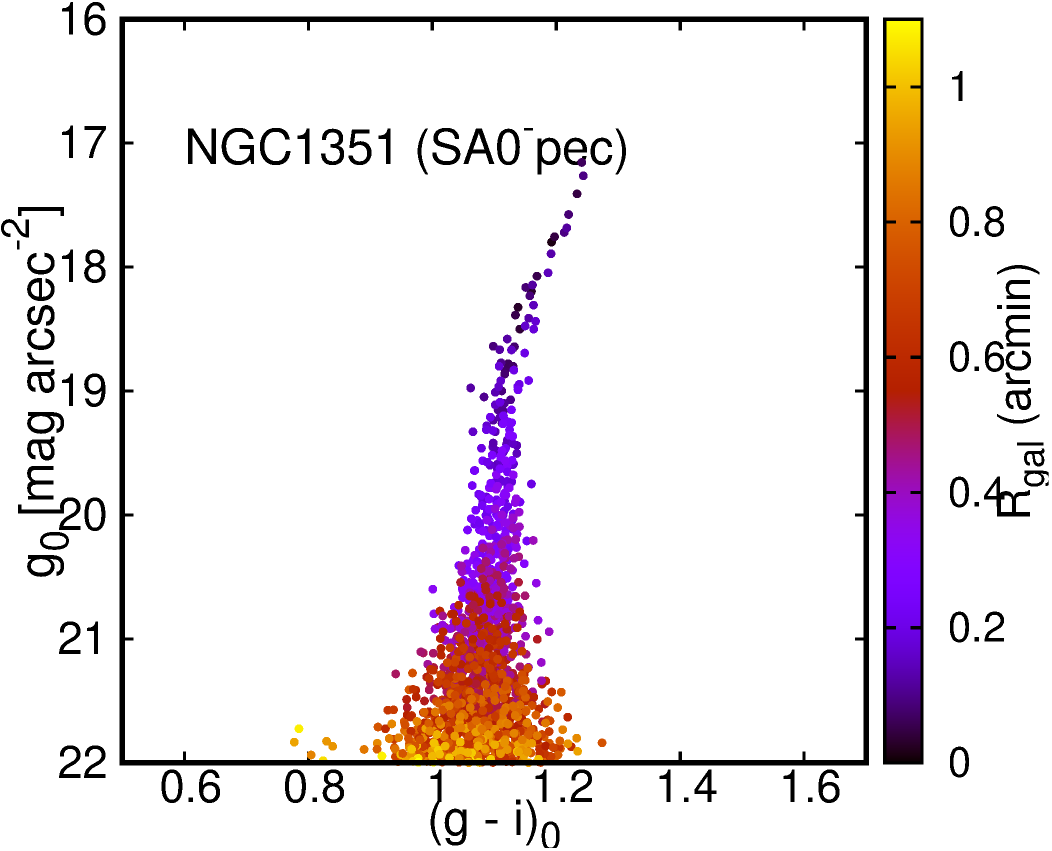}
\includegraphics[width=0.5\columnwidth]{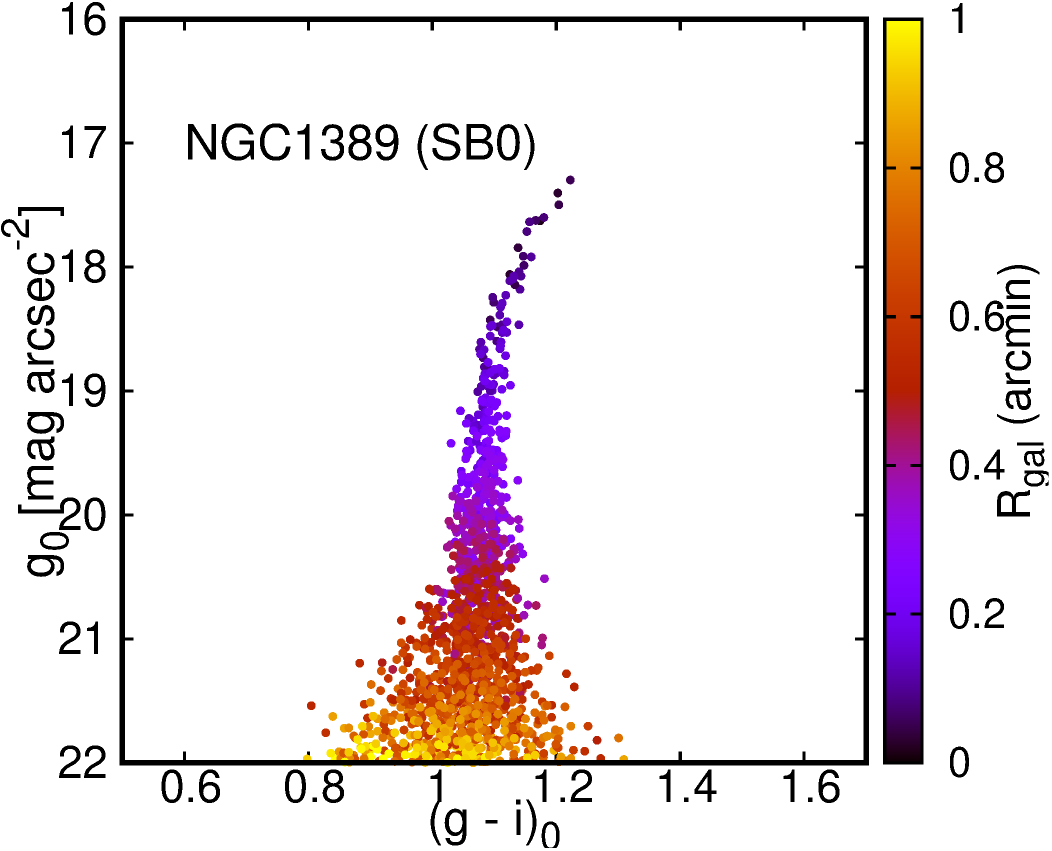}
\caption{Pixel color-magnitude diagrams of the analyzed lenticular galaxies, organized by decreasing luminosity (from left to right, and from top to bottom). The morphological type is indicated between parentheses. The color bar gives the galactocentric distance code for each pixel.
}
\label{Figure2}
\end{figure*}

\begin{figure}[!t]
\centering
\includegraphics[width=0.49\columnwidth]{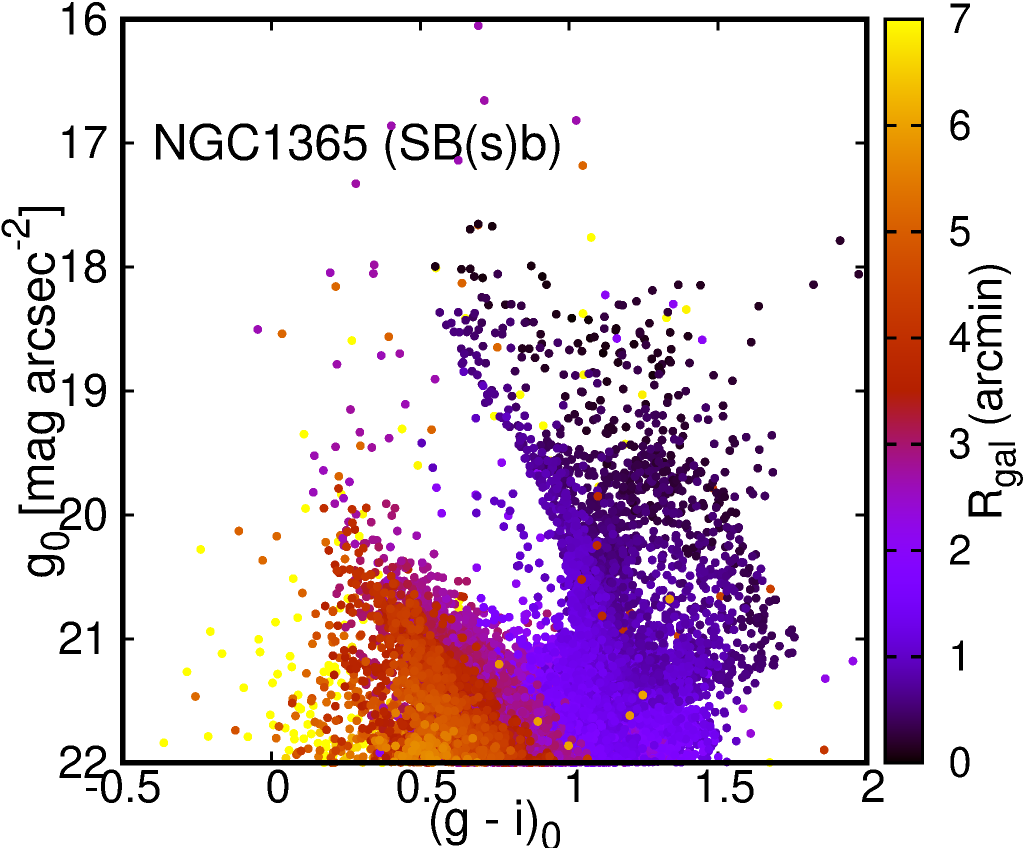}
\includegraphics[width=0.49\columnwidth]{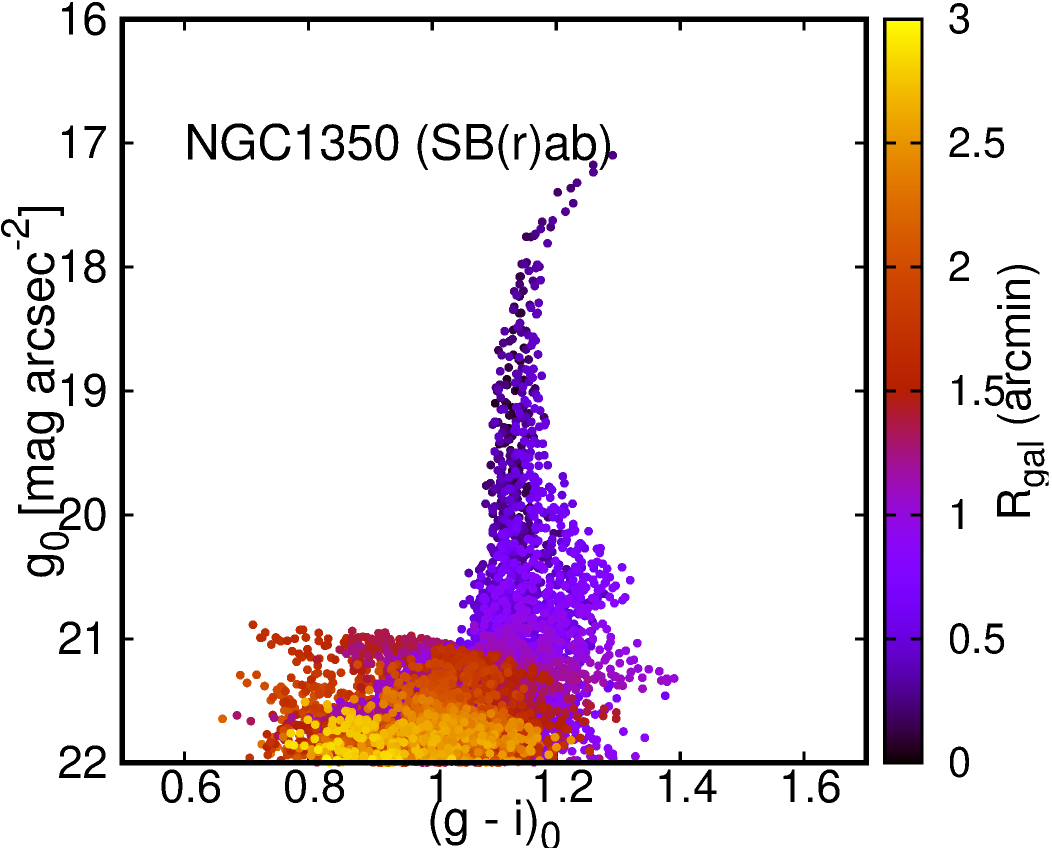}
\includegraphics[width=0.49\columnwidth]{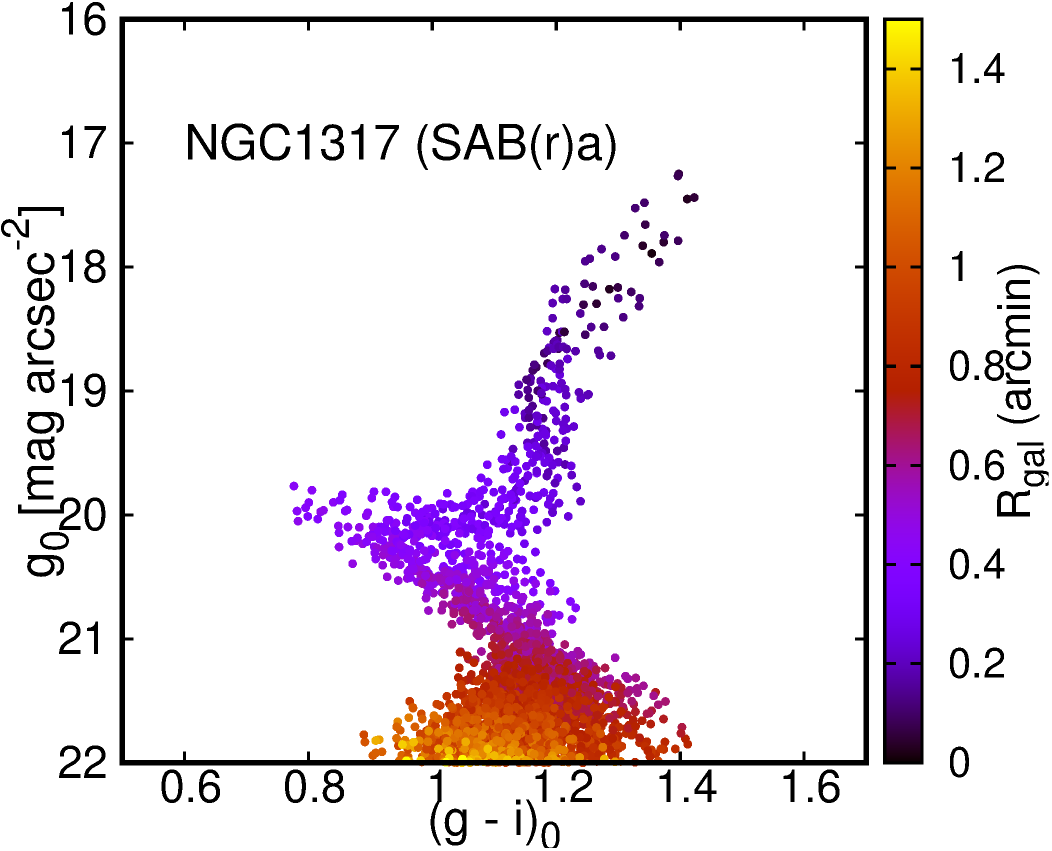}
\includegraphics[width=0.49\columnwidth]{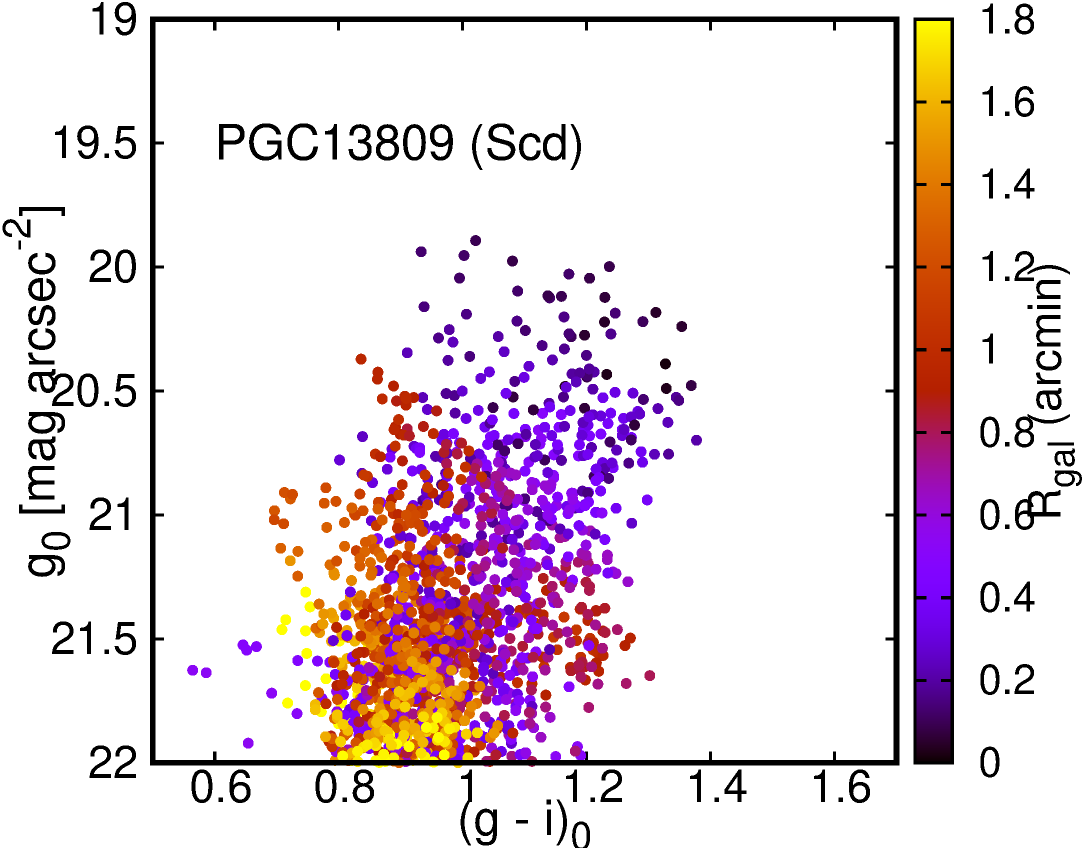}
\caption{Pixel color-magnitude diagrams of the analyzed spiral galaxies, organized by decreasing luminosity (from left to right, and from top to bottom). The morphological type is indicated between parentheses. The colored bar indicates the galactocentric distance of the pixels considered in each galaxy.
}
\label{Figure3}
\end{figure}

Although the sample analyzed in this work includes 24 galaxies, here we will only present a small group according to morphological type, in order to show the different characteristics among them. In this work, we consider galaxies with M$_g<-18$ mag in order to obtain a significant number of pixels with S/N$>$5 for each of them. Figures \ref{Figure1}, \ref{Figure2} and \ref{Figure3} show the pCMDs for elliptical (E), lenticular (S0) and spiral (S) galaxies, respectively, color-coded according to the distance of each pixel from the galactic center.

At first glance, pCMDs associated to E galaxies (Figure \ref{Figure1}) show the same behavior, defining a clear main sequence with average colors $1.1<(g-i)_0<1.2$ mag\,arcsec$^{-2}$. This is expected for the E due to their regular structures and the homogeneity in their ages. Furthermore, it is observed that the brightest pixels have slightly redder colors ($(g-i)_0 > 1.2$ mag\,arcsec$^{-2}$) compared to the aforementioned sequence. 

\begin{figure}[!t]
\centering
\includegraphics[width=0.50\columnwidth]{Figura1c.pdf}
\includegraphics[width=0.42\columnwidth]{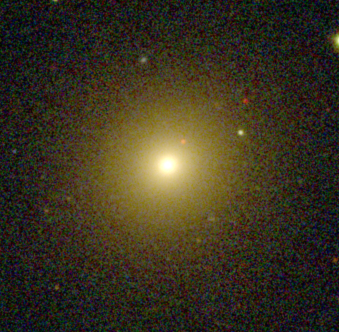}
\includegraphics[width=0.50\columnwidth]{Figura2f.pdf}
\includegraphics[width=0.42\columnwidth]{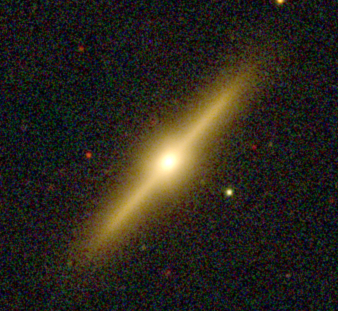}
\caption{\emph{Top left panel:} pixel color-magnitude diagram of the galaxy NGC\,1374 shown in the \emph{top right panel}. \emph{Bottom left panel:} pixel color-magnitude diagram of the galaxy NGC\,1381 shown in the \emph{bottom right panel}. Due to the similarities between the pCMDs of both galaxies, we wonder if NGC\,1374 could be a face-on S0 misclassified as E as a consequence of an inclination effect. }
\label{Figure4}
\end{figure}

On the other hand, although most of the diagrams of the S0 galaxies (Figure \ref{Figure2}) resemble that of the E ones, some of them show complex shapes, such as NGC\,1316, NGC\,1326, NGC\,1386, among others. The different shapes in these diagrams are mainly due to the presence of dust, structures such as bulges, discs, rings, or even star forming regions. It is worth noticing that the pCMDs of NGC\,1374, classified as E0, and NGC\,1381, classified as S0, are quite similar. We wonder if both galaxies are of the same morphological type, and one of them is misclassified due to a projection/inclination effect. For example, while NGC\,1381 is a clear edge-on S0 galaxy, NGC\,1374 looks like an elliptical galaxy but it could also be a face-on S0 (Figure \ref{Figure4}). Although, in principle, the analysis of its surface brightness profile would allow disentangling its different components, thus being able to determine its morphological type in some cases this analysis is complex and not conclusive. 

Finally, in the case of S galaxies (Figure \ref{Figure3}), it is observed that their diagrams show more complex structures compared to those of early type galaxies. The pCMDs of spiral galaxies present bluer colors in their outermost pixels ($(g-i)_0<0.9$ mag\,arcsec$^{-2}$) with respect to their central regions ($(g-i)_0\sim1.2$ mag\,arcsec$^{-2}$) and a wider color dispersion. This is due to the presence of star-forming regions, generally associated with spiral arms. Another feature observed in these diagrams is that the color dispersion is influenced by the inclination of the galaxy with respect to the sky plane.

\section{Conclusions}\label{sec:conclusion}
Using photometric data from the S-PLUS survey, we built individual pCMDs for 24 galaxies located in the Fornax cluster. By comparing the different diagrams grouped according to their morphological types, certain trends and differences among them are identified. For example, from the particular cases of NGC\,1374 and NGC\,1381, we wonder if this kind of diagrams could become a powerful tool to clarify dubious morphological classifications, besides allowing the identification of internal substructures.

As future work, we plan to obtain as many pCMDs as possible in Fornax, in order to have a significant statistical sample that will allow us to disentangle the issues mentioned above. In addition, we will explore the ability of these diagrams to provide information on stellar populations in Fornax \citep[e.g.,][]{Conroy2016,Lee2018}.

\hspace{0.5cm}

\begin{acknowledgement}
S-PLUS is an international collaboration founded by Universidade de Sao Paulo, Observatório Nacional, Universidade Federal de Sergipe, Universidad de La Serena and Universidade Federal de Santa Catarina. This work was funded with grants from Consejo Nacional de Investigaciones Científicas y Técnicas de la República Argentina, and Universidad Nacional de La Plata (Argentina).
\end{acknowledgement}


\bibliographystyle{baaa}
\small
\bibliography{bibliografia}
 
\end{document}